# CREATE: Cohort Retrieval Enhanced by Analysis of Text from Electronic Health Records using OMOP Common Data Model

Sijia Liu[a,b], Yanshan Wang[a], Andrew Wen[a], Liwei Wang[a], Na Hong[a], Feichen Shen[a], Steven Bedrick[c], William Hersh[d], Hongfang Liu[a]

[a] Department of Health Sciences Research, Mayo Clinic, Rochester, MN, USA

[b] Department of Computer Science and Engineering, University at Buffalo, The State University of New York, Buffalo, NY, USA

[c] Department of Computer Science & Electrical Engineering, Oregon Health & Science University, Portland, OR, USA

[d] Department of Medical Informatics & Clinical Epidemiology, Oregon Health & Science University, Portland, OR, USA

**Abstract** (150-250 words)

*Background*: Widespread adoption of electronic health records (EHRs) has enabled secondary use of EHR data for clinical research and healthcare delivery. Natural language processing (NLP) techniques have shown promise in their capability to extract the embedded information in unstructured clinical data, and information retrieval (IR) techniques provide flexible and scalable solutions that can augment the NLP systems for retrieving and ranking relevant records.

*Methods*: In this paper, we present the implementation of Cohort Retrieval Enhanced by Analysis of Text from EHRs (CREATE), a cohort retrieval system that can execute textual cohort selection queries on both structured and unstructured EHR data. CREATE is a proof-of-concept system that leverages a combination of structured queries and IR techniques on NLP results to improve cohort retrieval performance while adopting the Observational Medical Outcomes Partnership (OMOP) Common Data Model (CDM) to enhance model portability. The NLP component empowered by cTAKES is used to extract CDM concepts from textual queries. We design a hierarchical index in Elasticsearch to support CDM concept search utilizing IR techniques and frameworks.

*Results*: Our case study on 5 cohort identification queries evaluated using the IR metric, P@5 (Precision at 5) at both the patient-level and document-level, demonstrates that CREATE achieves an average P@5 of 0.90, which outperforms systems using only structured data or only unstructured data with average P@5s of 0.54 and 0.74, respectively.

**Keywords (4-6)**: cohort retrieval, information retrieval, common data model, electronic health records, natural language processing

# 1 Introduction

The widespread adoption of Electronic Health Records (EHRs) has enabled secondary use of EHR data for clinical research and healthcare delivery [1]. Many institutions have established clinical data repositories (CDRs) in conjunction with cohort discovery tools, such as i2b2, to enable investigators to use EHR data for cohort identification in clinical trials and retrospective clinical studies. As much of the detailed patient information is embedded in clinical narratives, cohort identification using only structured data such as diagnosis codes or procedure codes has limited retrieval performance [2–5]. Natural Language Processing (NLP) techniques have shown promise to be leveraged for various applications in clinical research [6–9]. There are many existing clinical NLP systems developed to encode information from unstructured EHR data [10, 11]. Successful applications of clinical NLP to translational research in phenotyping, clinical workflow optimization [6, 12] and quality control [13, 14] have been reported to facilitate both clinical research and pharmacogenomics studies.

Information retrieval (IR) techniques, which retrieve and rank documents from a large collection of text documents based on users' queries, can provide an alternative approach for leveraging clinical narratives for cohort identification [15, 16]. As the performance of existing clinical NLP systems for concept encoding is still unsatisfactory [17, 18], the combination of NLP and IR is a promising solution for EHR-based cohort retrieval.

Clinical Common Data Models (CDMs) are designed to provide a standardized and logically unified way to represent EHR data from distributed research networks. CDMs ensure that clinical research methods are consistent and reusable across the networks for producing meaningful, comparable and reproducible results [19, 20]. Multiple CDMs exist to support large-scale research networks such as ACT [1], eMERGE [21], and PCORnet [22]. Our prior investigation demonstrated the generalizability of the Observational Medical Outcomes Partnership (OMOP) CDM by the Observational Health Data Sciences and Informatics (OHDSI) Program [23] in multi-institutional research [24]. CDMs have the ability to achieve both structural and semantic consistence of EHR data in clinical data research networks, but it is still an open question as to how CDMs can be utilized to represent textual cohort criteria or queries.

Moving beyond the current CDR system design and implementation, an efficient and comprehensive patient-level search engine on both structured and unstructured EHR data is therefore still highly demanded by healthcare practitioners and researchers. In this paper, we describe an end-to-end cohort retrieval system named Cohort Retrieval Enhanced by Analysis of Text from EHRs (CREATE) using the OMOP CDM. Cohort retrieval in CERATE is conducted in two phases: the first phase filters patients using structured data, and the second phase retrieves and ranks results at either a document or a patient level. The system was tested using a query collection assembled previously [25] on a corpus composed of the EHR data from the Mayo Biobank cohort [26].

---

[1] http://www.act-network.org/

## 2 Methods

### 2.1 Overview of system architecture

An overview of our proposed cohort retrieval system in CDR is shown in Figure 1. Specifically, a textual query will be expanded and divided, either automatically or manually, into structured and unstructured data fields according to specific CDR implementations. The query fulfillment for structured data and text data will be managed differently. Structured EHR data can be retrieved from the corresponding CDRs using Structured Query Language (SQL) on a relational database management system (RDBMS) and the unstructured EHR data can be pre-processed by NLP and retrieved leveraging IR techniques [27]. Retrieved results can then be combined and aggregated for clinical research applications, such as clinical trial feasibility assessments or cohort identification. The retrieved screened cohort can be treated as a weakly labeled dataset for cohort identification. A potential following step is human relevance judgment to manually validate the retrieved results through chart review.

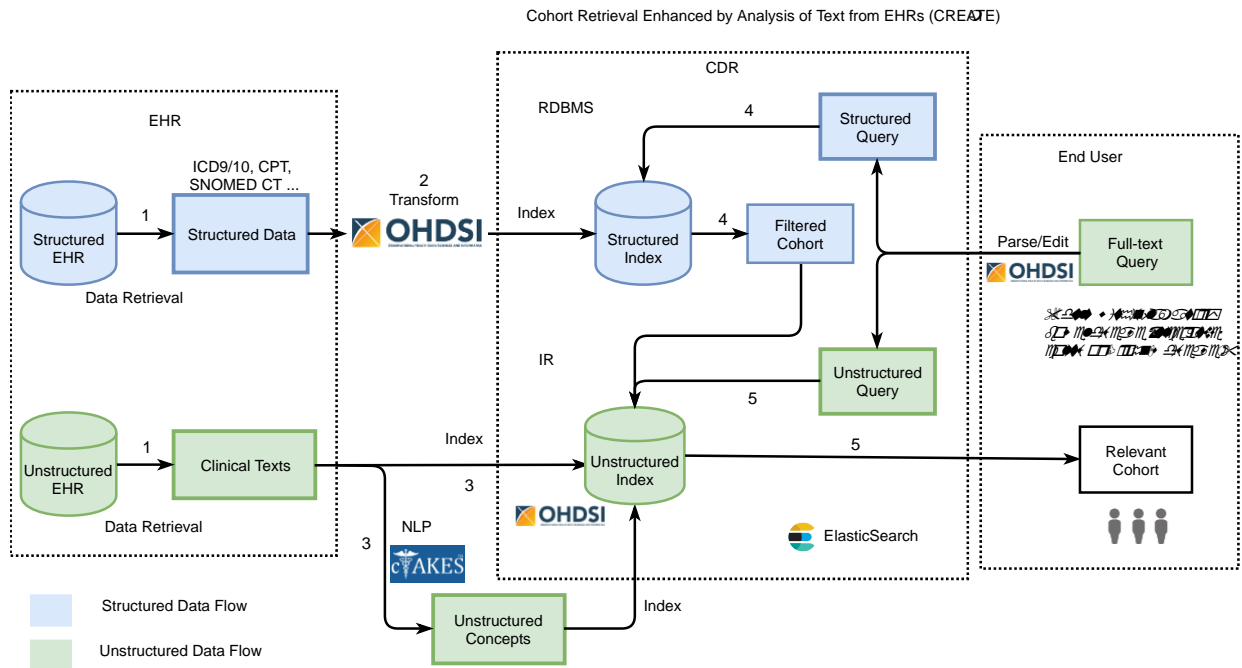

Figure 1: Overview of CREATE's workflow: 1) Data retrieval: retrieving both the structured and unstructured EHR data that will be used for cohort retrieval tasks. 2) CDM transformation of structured data: transforming raw structured data into CDM vocabularies using mapping tables or NLP. 3) CDM transformation of unstructured data using NLP: extracting CDM concepts from the clinical narratives using information extraction software and building an index using these extracted concepts. 4) Filtering querying cohort using CDM concepts sourced from structured data: using structured fields in the queries to reduce the potential patient candidates for IR. 5) IR using CDM concepts: retrieving and ranking on the screened patients to query the unstructured fields.

## 2.2 Adopting OMOP CDM for patient retrieval

To improve the interoperability and portability of our system with disparate data sources, we adopt the OMOP Common Data Model V5.0[2] as the data model to index EHR data. The hierarchical index structure of CDRs using OMOP CDM for cohort retrieval is shown in Figure 2. The indexed tables include data from both unstructured and structured sources, consisting of extracted OMOP CDM artifacts from unstructured clinical notes, as well as encounter information, demographic information (represented as a CDM Person), and diagnoses, procedures, and lab tests from structured data. The distinction between structured and unstructured data may vary between different EHR systems. The specifics of implementation in adopters may therefore differ slightly from our implementation in this study.

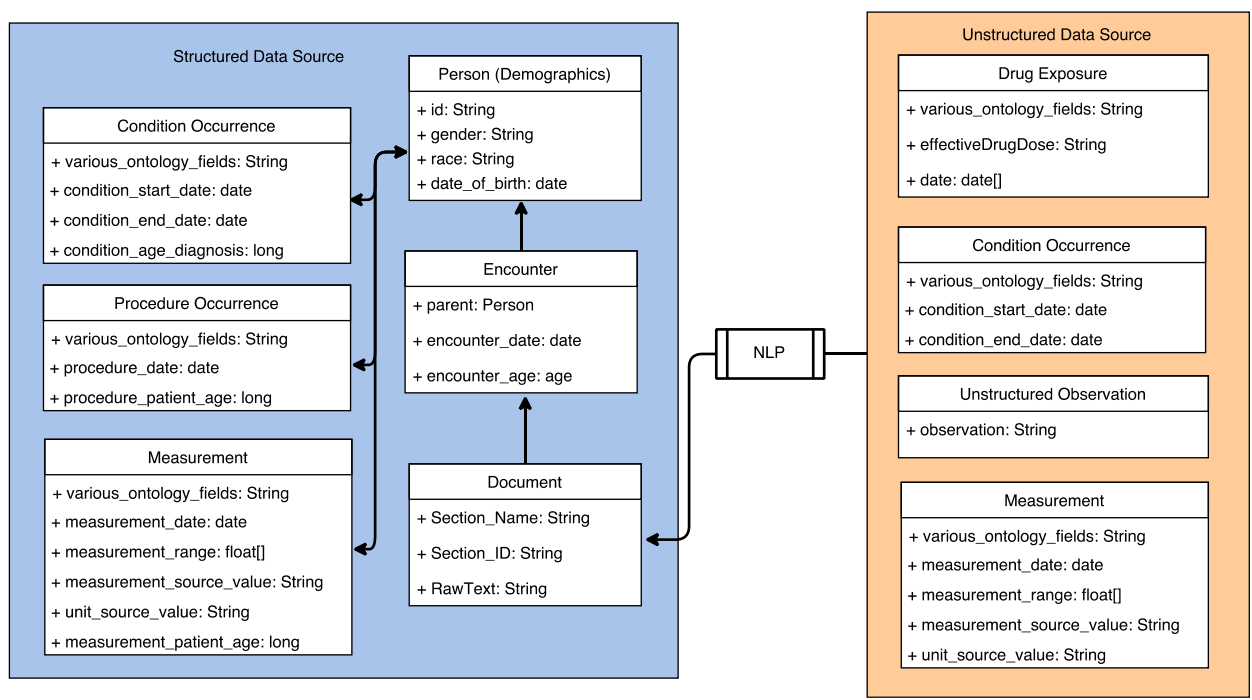

Figure 2 Hierarchical index structure using OMOP CDM. The "various_ontology_fields" include objects such as the OHDSI vocabulary code, UMLS CUIs, SNOMED-CT codes, Rx_Norm and corresponding normalized texts

### 2.2.1 Structured data

Structured data such as procedures, diagnosis, lab tests and demographics are directly queried from relational databases and loaded into the index through an ETL process. We map structured data to UMLS concept unique identifiers either through the usage of mapping definitions already in the UMLS Metathesaurus[3] (e.g., ICD9/10, CPT, and SNOMEDCT) or through the use of NLP (e.g., local lab test codes). The concepts are subsequently mapped to equivalent OHDSI/OMOP compliant vocabulary codes via ATHENA standardized vocabularies[4].

---

[2] http://www.ohdsi.org/web/wiki/doku.php?id=documentation:cdm:single-page

[3] https://www.nlm.nih.gov/research/umls/knowledge_sources/metathesaurus/

### 2.2.2 Unstructured data

A typical clinical document consists of multiple sections that each provides an essential yet brief description of a specific perspective from a patient encounter, such as the patient's social history, diagnosis or chief complaints. We choose to use document sections as the unit to index for cohort retrieval based on the observation that while retrieval at a sentence level is insufficient for relevance judgment purposes in the topic collections we investigated, document level retrieval of each encounter provides too much extraneous, mostly irrelevant, information, with descriptions reaching several pages in length.

Various CDM concepts are extracted via NLP from individual clinical documents and subsequently indexed into Elasticsearch. Specifically, these concepts are extracted using the Apache cTAKES clinical NLP pipeline. Additionally, NLP concept attributes such as negation, certainty and family history are stored in the field "term_modifiers".

### 2.3 Textual query modeling

Textual queries in natural language are fed into the same NLP component used for indexing. Similarly, the normalized concepts and their associated attributes (e.g. negation, certainty, experiencer, status) are extracted from the textual query. Logical concepts such as "must" and "must not" are also used when generating queries from text for further parsing and interpretation in the query backend. An example of the textual query modeling process is illustrated in Figure 3. In the query "*Adults with inflammatory bowel disease (ulcerative colitis or Crohn's disease), who have not had surgery of the intestines, rectum, or anus entailing excision, ostomy*", the NLP component can detect and normalize the raw mentions of "bowel disease", "ulcerative colitis" and "Crohn's disease" into various coding systems including OHDSI IDs, while the demographic information of "adults" and the list of surgeries can be manually added as structured data filters based on date of birth and CPT codes. All the extracted information is shown to the users and subject to manual review and modification before query execution.

---

[4] https://www.ohdsi.org/analytic-tools/athena-standardized-vocabularies/

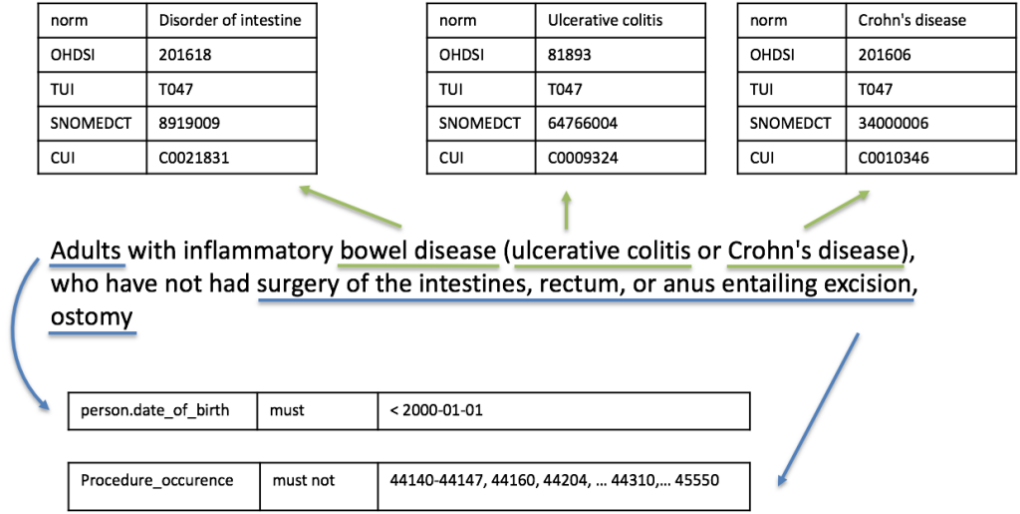

Figure 3 Textual query modeling of the query "Adults with inflammatory bowel disease (ulcerative colitis or Crohn's disease), who have not had surgery of the intestines, rectum, or anus entailing excision, ostomy".

### 2.4 Retrieval methods

CREATE uses Elasticsearch[5] as the search engine of the backend IR component. Since Elasticsearch includes support for hierarchical queries of parent/child relations, the hierarchical index architecture allows for significant flexibility in query strategies. For instance, individual documents with a certain set of CDM attributes can be discovered, but those documents can additionally be filtered at query time by encounter age and whether the patient they are associated with also has documents with other attributes and/or has a certain set of structured data. Similarly, this index structure allows for other variations of search objects. The search is not limited to documents, but also applies to patients with a certain set of attributes within their clinical notes and structured data.

Given a document $d$ and a textual query $q$, the set of CDM concepts extracted from $q$ can be represented as $O = \{o_1, o_2, ..., o_M\}$, where $o$ is a CDM concept. The similarity score between $d$ and $o$ can then be represented as $s(d, o)$. The total score of each document for each query $s_q$ would then be defined as:

$$s_q(d) = \frac{1}{M} \sum_{o \in O} s(d, o) + s(d, q)$$

The first term on the right-hand side of the equation is the average similarity of all CDM concepts in the query. The second term is the similarity between the document and the full-text query. In extreme use cases, the two terms can be weighted to place more emphasis on the contribution of either structured or unstructured data to the query. The patient-level similarity score is the average of the top 100 document scores. The top rank threshold 100 is selected based on our experiments on top 10/20/50/100 from test query results and may subject to further tuning based on EHR systems.

---

[5] https://www.elastic.co/guide/en/elasticsearch/reference/current/index-modules-similarity.html

## 3 Experiments

We implemented CREATE as a study feasibility assessment tool for Mayo Biobank Rochester cohort, which is a large-scale institutionally funded research resource initiated in 2009 with blood, EHR and patient provided data on 45,613 Mayo Clinic Rochester patients who had consented to participate regardless of their health history. This resource has been used for over 250 research studies in a wide array of health related research and clinical studies [26]. In our experiments, we limited included patients to patients with at least one clinical note in our EHR and extracted the corresponding structured EHR for those patients.

After the data extraction from Mayo EHR, we investigated and compared the EHR system implementation at Mayo Clinic to OMOP CDM tables. During the data exploration stage, we found that the data elements under those corresponding tables were generally straightforward to map. Therefore, we just showed the mapping at the granularity of table level. Table I shows our mapping of several OMOP CDM tables to Mayo EHR tables. cTAKES was used as our NLP system, and the mapping used to transform named entity mention types of the cTAKES type system to CDM tables is also listed in Table I. We have developed a web-based user interface for CREATE, the details of which are described in Appendix A.

Table I Table-level mapping between OMOP CDM and Mayo EHR

| OMOP CDM Table | Mayo CDR | Number of records | Vocabulary | NLP - cTAKES type system |
|---|---|---|---|---|
| Person | Demographics | 45,613 | - | |
| Condition | Diagnosis | 9,712,736 | ICD9, ICD10 | SignSymptom DiseaseDisorder |
| | Procedures | 13,014,264 | CPT | Procedure |
| Measurement | Lab | 15,719,203 | ROCLIS | Lab |
| | Vital Signs | - | - | VitalSigns |
| Drug_exposure | DrugExposure | - | UMLS | Medication |
| Unstructured Conditions | Clinical notes | 68,198,499 | - | - |

As an evaluation of CREATE, we randomly sampled 5 queries from a previously curated query collection [25] to evaluate CREATE through manual chart review. Table II lists the detailed description of the queries and the corresponding key words used in the manual chart review process for judging patient relevance. Note that the queries are significantly different from the single condition criteria used to evaluate systems in related works in regards to the level of detail, logic, and semantic complexity involved. The complete parsing results of the structured part of the queries can be found in Appendix B, followed by the CREATE query format specification (Appendix C).

Table II The list of tested queries

| Query | Description | Key words |
|---|---|---|
| 1 | Adults with inflammatory bowel disease (ulcerative colitis or Crohn's disease), who have not had surgery of the intestines, rectum, or anus entailing excision, ostomy | ulcerative colitis, Crohn's disease, excision, ostomy, rectal prolaspse, anal fistula, stricturoplasty resection |
| 2 | Adults 18-100 years old who have a diagnosis of hereditary hemorrhagic telangiectasia (HHT), which is also called Osler-Weber-Rendu syndrome. | Osler-Weber-Rendu syndrome, hereditary hemorrhagic telangiectasia |
| 3 | Children with localization-related (focal) epilepsy with simple or complex partial seizures diagnosed before 4 years old who have had an outpatient neurology visit. | Epilepsy, partial seizure, neurology, ACE |
| 4 | Adults 18-70 years old with rheumatoid arthritis currently treated with methotrexate who have never used a biologic disease-modifying antirheumatic drug (DMARD). | rheumatoid arthritis biologic methotrexate abatacept, adalimumab, anakinra, certolizumab, etanercept, golimumab, infliximab, rituximab, tocilizumab, tofacitinib |
| 5 | Adults who have been treated with an angiotensin-converting-enzyme (ACE) inhibitor and developed an associated cough, consistent with ACE inhibitor-induced cough as an adverse effect of the medication. | Benazepril, Lotensin, Captopril, Enalapril, Vasotec, Fosinopril, Lisinopril, Prinivil, Zestril, Moexipril, Perindopril, Aceon, Quinapril, Accupril, Ramipril, Altace, Trandolapril, Mavik, cough, angiotensin-converting-enzyme (ACE) inhibitor |

Performance was measured using the average P@5 of the 5 queries. The structured query used the manually transformed ICD-9/10 codes. There was no ranking of relevance for the retrieved patients from structured EHR data, thus we randomly selected 5 patients from the relevant patients to calculate the P@5. The top 5 patients from unstructured EHR query and CREATE results were retrieved based on BM25 [28]. A medical expert performed complete chart review on the top 5 patients for each retrieval cohort. The patient relevancy was scored into the three categories "Definitely Relevant" (DR), "Partially Relevant" (PR) and "Not Relevant" (NR) by the medical expert. DR, PR, and NR were assigned to scores of 1, 0.5 and 0, respectively, for P@5 calculation.

Table III P@5 of sampled queries

| Query | Results on structured EHR data | Results on unstructured EHR data | CREATE (combined) |
|---|---|---|---|
| 1 | 0.8 | 0.6 | 0.8 |
| 2 | 0.7 | 1.0 | 1.0 |
| 3 | 0.3 | 0.5 | 0.8 |
| 4 | 0.7 | 0.7 | 1.0 |
| 5 | 0.2 | 0.9 | 0.9 |
| Average | 0.54 | 0.74 | 0.90 |

The result P@5s are shown in Table III. The overall comparison shows that CREATE, as a combination of systems using structured and unstructured EHR data, outperformed the systems based on using only one of structured or unstructured EHR data for full-text queries. For each query, CREATE performs at least as well as the systems using only structured or unstructured EHR data.

## 4  Discussion

The proposed system CREATE is in essence a proof-of-concept for leveraging the combination of structured queries and IR techniques to improve cohort retrieval performance while adopting the OMOP CDM to enhance model portability. The implementation and evaluation using sample queries on our cohort support our hypothesis that using a combination of structured and unstructured EHR data would outperform the single-source system in determining the relevance between a given patient and the input query. CREATE aims to improve the efficiency of judging patient relevance in the EHR, by shifting from human-query judgment (pull) to system-feed judgment (push). The automatic term extraction and normalization significantly reduces the amount of time needed for future manual mapping, since most of the concepts will be correct and only require the action of verification instead of wild searching.

### 4.1 Related Works

There are generally two approaches to search unstructured EHR data for purposes such as patient care, clinical research, and traceability of medical care [29]. The first approach is based on text search. For example, the Electronic Medical Record Search Engine (EMERSE) from the University of Michigan [15] is a full-text search engine based on Apache Lucene[6] with the goal of facilitating the retrieval of information for clinicians, administrators, and clinical and translational researchers based on clinical narratives. A key drawback, however, is that it does not support queries using structured EHR data such as demographic information, lab tests and medications. Dr. Warehouse proposed by Garcelon et al [30] is a free-text search engine using Oracle Text to index all its documents. The system is based on relational databases and relies on ranking after retrieval which may limit its capability to deploy state-of-the-art IR methods such as BM25 or Markov Random Fields. The other approach to searching unstructured EHR data is to extract concepts in unstructured EHR data using NLP systems. For example, SemEHR [31] is a semantic search engine based on a Fast Healthcare Interoperability Resources (FHIR)[7] representation of the extracted clinical semantic concepts from a clinical NLP system, Bio-YODIE[8]. The system showed a high performance in retrieving patients given queries of single concepts such as Hepatitis C and HIV in local EHR and lab tests measurements when evaluated against the MIMIC-III dataset.

In comparison with existing search engines in unstructured EHR data, our cohort retrieval system has the following characteristics: 1) the adoption of CDMs to facilitate cohort retrieval using both structured and unstructured data for multi-institutional research; 2) the flexibility and ability to apply state-of-the-art IR methods in the retrieval system; 3) the incorporation of relevance judgment for downstream machine learning based cohort selection methods; and 4) the generation of semantic annotations during the indexing phrase to provide a real-time semantic search experience.

### 4.2 Limitations

This study has multiple limitations that may offer directions for our future work:

Our current evaluation is based on P@5, which is a relatively small evaluation set for general cohort retrieval tasks. Though we acknowledge that a larger evaluation on a fully-annotated patient cohort would be helpful to better evaluate the proposed system, it is time-consuming to do judgment on the complete Biobank dataset.

When processing concepts without a global coding system like CPT or ICD9/10, the concept mapping in our current solution relies on NLP algorithms. Although it is a fast and straightforward solution, the current NLP tools cannot achieve the same level of accuracy as human assigned codes. Complete mapping from a local vocabulary requires extensive human efforts with data quality assurance [32], thus it is not feasible within the scope of this study. A solution for this issue is to utilize value set repositories to manage the concepts. Though a one-to-one mapping may not be found in all semantic spaces, value set repositories can provide a systematic way to manage the concept sets in collections or aggregations [33].

---

[6] https://lucene.apache.org/

[7] https://www.hl7.org/fhir/resourcelist.html

[8] https://gate.ac.uk/applications/bio-yodie.html

There are also several potential approaches to further improve the IR component in the current framework. We only used the out-of-box query algorithms implemented by Elasticsearch to measure the patient similarity and rank the relevancy in this study. More advanced IR methods can be applied to the queries such as case-based reasoning [34–36], pseudo relevance feedback [37] and different ranking models [38, 39]. Though the equal weights of CDM concepts and raw text provide information from both sides, the weights can be tuned to meet different retrieval perspectives and demands.

## 5 Conclusion

We developed CREATE, an end-to-end patient-level IR system, with the ability to query both structured and unstructured data leveraging the OMOP Common Data Model. The implementation and evaluation on Mayo Clinic Biobank demonstrated that CREATE outperforms cohort retrieval systems using only one of either structured or unstructured data in complex textual cohort queries.

In the future, we will refine the evaluation process by adding more topics and larger cohort of manual chart reviews. An active learning component will be added to the system to enable Human-In-The-Loop analysis on the screened cohort from the system to further improve the efficiency of relevance judgment. By doing so, both machine learning-based or rule-based cohort identification algorithms could be deployed and evaluated in real-time. This could potentially then be extended to an active learning cohort identification framework [40] once integrated with machine learning models for cohort identification.

## Acknowledgments

We sincerely thank our annotator Donna Ihrke, who annotated the query corpus. The work was supported by National Institutes of Health grants R01LM011934, R01EB19403, R01LM11829, and U01TR02062. The content is solely responsibility of the authors and does not necessarily represent the official views of the National Institutes of Health.